\documentclass[twocolumn,preprintnumbers,amsmath,amssymb,nobibnotes,nofootinbib,superscriptaddress]{revtex4}
\usepackage{graphicx}% Include figure files
\usepackage{dcolumn}% Align table columns on decimal point
\usepackage[usenames,dvipsnames]{color}
\RequirePackage[colorlinks=true
,urlcolor=blue
,anchorcolor=blue
,citecolor=blue
,filecolor=blue
,linkcolor=blue
,menucolor=blue
,pagecolor=blue
,linktocpage=true
,pdfproducer=medialab
]{hyperref}
\usepackage{amsmath}
\usepackage{mathrsfs}
\usepackage{slashed}
\usepackage{cancel}

\newcommand{\LL}{\mathscr{L}}

\def\cD{{\cal D}}
\def\cF{{\cal F}}
\def\cO{{\cal O}}
\def\cP{{\cal P}}

\def\cY{{\bf Y}}

\def\Tr{{\rm Tr}}

\def\be{\begin{equation}}
\def\ee{\end{equation}}
\def\beq{\begin{equation}}
\def\eeq{\end{equation}}
\def\bc{\begin{center}}
\def\ec{\end{center}}
\def\bea{\begin{eqnarray}}
\def\eea{\end{eqnarray}}

\def\nn{\nonumber}

\newcommand{\mean}[1]{\langle#1\rangle}
\newcommand{\derp}{\partial}

\newcommand{\diag}{{\rm{\bf diag}}}
\newcommand{\UH}{\mathbf{U}}

\newcommand{\TL}{\mathbf{T}}
\newcommand{\VL}{\mathbf{V}}
\newcommand{\DL}{\mathbf{D}}
\newcommand{\WL}{\mathbf{W}}

\newcommand{\red}[1]{\color{red} #1 \color{black}}
\newcommand{\blue}[1]{\color{blue} #1 \color{black}}
\newcommand{\green}[1]{\color{OliveGreen} #1 \color{black}}

\begin{document}
%
%%%
%%%%%%%%%%%%%%%%%%% Title Page%%%%%%%%%%%%%%%%%%%%%%%%%
%%%
%

\preprint{FTUAM-12-115}
\preprint{IFT-UAM/CSIC-12-113}
\preprint{CERN-PH-TH/2012-335}
\preprint{DFPD-2012/TH/23}

\title{The Effective Chiral Lagrangian for a Light Dynamical ``Higgs Particle''}

\author{ R. Alonso}
\email{rodrigo.alonso@uam.es}
\affiliation{Departamento de F\'isica Te\'orica and Instituto de F\'{\i}sica Te\'orica, IFT-UAM/CSIC,\\
Universidad Aut\'onoma de Madrid, Cantoblanco, 28049 Madrid, Spain}

\author{M. B. Gavela}
\email{belen.gavela@uam.es}
\affiliation{Departamento de F\'isica Te\'orica and Instituto de F\'{\i}sica Te\'orica, IFT-UAM/CSIC,\\
Universidad Aut\'onoma de Madrid, Cantoblanco, 28049 Madrid, Spain}
\affiliation{CERN, Department of Physics, Theory Division CH-1211 Geneva 23, Switzerland}

\author{L. Merlo}
\email{luca.merlo@uam.es}
\affiliation{Departamento de F\'isica Te\'orica and Instituto de F\'{\i}sica Te\'orica, IFT-UAM/CSIC,\\
Universidad Aut\'onoma de Madrid, Cantoblanco, 28049 Madrid, Spain}
\affiliation{CERN, Department of Physics, Theory Division CH-1211 Geneva 23, Switzerland}

\author{S. Rigolin}
\email{stefano.rigolin@pd.infn.it}
\affiliation{Dipartimento di Fisica ``G.~Galilei'', Universit\`a di Padova and \\
INFN, Sezione di Padova, Via Marzolo~8, I-35131 Padua, Italy}

\author{J. Yepes}
\email{ju.yepes@uam.es}
\affiliation{Departamento de F\'isica Te\'orica and Instituto de F\'{\i}sica Te\'orica, IFT-UAM/CSIC,\\
Universidad Aut\'onoma de Madrid, Cantoblanco, 28049 Madrid, Spain}

\begin{abstract}
We generalize the basis of CP-even chiral effective operators describing a dynamical Higgs sector, to the 
case in which the Higgs-like particle is light. Gauge and gauge-Higgs operators are considered up to four derivatives.
This analysis completes the tool needed to explore at leading order the connection 
between linear realizations of the electroweak symmetry breaking mechanism - whose extreme case is 
the Standard Model - and non-linear realizations with a light Higgs-like particle present. It may 
also provide a model-independent guideline to explore which exotic gauge-Higgs couplings may be expected, 
and their relative strength to Higgsless observable amplitudes. With respect to fermions, the analysis is reduced by nature to the consideration of those flavour-conserving operators that can be written in terms of pure-gauge or gauge-Higgs ones via the equations of motion, but for the standard Yukawa-type couplings.
\end{abstract}
\maketitle

%
%%%%%%%%%%%%%%%%%%%%%%%%%   1.  Introduction       %%%%%%%%%%%%%%%%%%%%%%%%
%
\section{Introduction}

A new resonance at the Electroweak (EW) scale has been established at LHC~\cite{:2012gk,:2012gu}, consistent 
with the hypothesis of the SM scalar boson (so-called ``Higgs boson'' for short hereafter)~\cite{Englert:1964et,
Higgs:1964ia,Higgs:1964pj} with mass around 125 GeV. 

There are essentially two main frameworks that have been proposed to explain the EW symmetry breaking sector. 
The first possibility is that the Higgs is a fundamental particle, transforming linearly (as a doublet in the 
standard minimal picture) under the gauge symmetry group $SU(2)_L \times U(1)_Y$.  Another possibility is, 
however, that the Higgs dynamics is not perturbative and the gauge symmetry in the scalar sector is non-linearly 
realized; this may be the case for instance if the Higgs resonance does not correspond to an elementary particle. 
In such a framework some strong dynamics should intervene at a scale $\Lambda_s$, and the characteristic scale 
of the associated Goldstone bosons $f$ respects $\Lambda_s \le 4\pi f$~\cite{Manohar:1983md}. In the original 
formulation~\cite{Susskind:1978ms,Dimopoulos:1979es,Dimopoulos:1981xc} the physical Higgs particle is simply 
removed from the low-energy spectrum and only the three would-be-Goldstone bosons are retained, in order 
to give masses to the weak gauge bosons, with $f=v$, where $v=246$ GeV denotes the electroweak scale defined 
via the $W$ mass, $M_W=g v/2$. The smoking gun signature of this ``technicolor" ansatz is the appearance 
of several vector and fermion resonances at the TeV scale. 
 
However, several variants of the strong interacting ansatz exist, with some of them ``predicting'' the existence 
of a light Higgs resonance in the spectrum. In the best known of such scenarios, originally proposed in 
Refs.~\cite{Kaplan:1983fs,Kaplan:1983sm,Banks:1984gj,Georgi:1984ef,Georgi:1984af,Dugan:1984hq}, the SM Higgs 
particle is substituted by a composite scalar degree of freedom that, being a quasi-Goldstone boson 
of a larger symmetry group, cannot acquire a large (i.e. $O$(TeV)) mass\footnote{See for example 
Ref.~\cite{Contino:2010rs} for a recent review on the subject.}. Besides this light Higgs-like scalar 
particle, these models still present a strongly interacting sector at the TeV scale, while they may correct 
at lower energies the size of SM couplings. This path looks promising in the absence of new resonances in 
LHC data. For these sophisticated constructions, the characteristic scale $f$ associated to the Goldstone 
bosons of the theory - which now include also the Higgs particle - does not need to coincide: i) neither with 
the scale of electroweak symmetry breaking, that will be denoted by $\mean{h}$, ii) nor with the electroweak 
scale $v$; while a constraint links together $f$, $\mean{h}$ and $v$. Indeed, in these hybrid schemes
\begin{equation}
\xi\equiv (v/f)^2
\label{xi}
\end{equation}
parametrizes  the degree of non-linearity of the Higgs dynamics. In the limit in which $\Lambda_s$ and 
thus $f$ go to infinity, the linear SM picture is recovered.

Without referring to a specific model, one can attempt to describe NP effects by making use of an effective 
Lagrangian approach, with operators made out of SM fields. The transformation properties of the longitudinal 
degrees of freedom of the electroweak gauge bosons can always be described at low-energy\footnote{Notice that in this low-energy expression for $\UH(x)$, the scale associated to the eaten GBs is $v$ and not $f$. Technically, the scale $v$ appears through a redefinition of the GB fields so as to have canonically normalized kinetic terms.} by a dimensionless unitary matrix transforming as a bi-doublet of the global symmetry group:
\beq
\UH(x)=e^{i\sigma_a \pi^a(x)/v}\, , \qquad \qquad  \UH(x) \rightarrow L\, \UH(x) R^\dagger\,,
\nn
\eeq
with $L,R$ denoting respectively the $SU(2)_{L,R}$ global transformations, spontaneously broken to the diagonal 
custodial  symmetry $SU(2)_C$, and explicitly broken by the $U(1)_Y$ gauge interaction and by the (different) 
masses of fermions in each $SU(2)_L$ fermion doublet. The adimensionality of $\UH(x)$ is the technical key to 
understand why the dimension of the leading low-energy operators describing the dynamics of the scalar 
sector differs for a non-linear Higgs sector~\cite{Appelquist:1980vg,Longhitano:1980iz,Longhitano:1980tm,
Feruglio:1992wf,Appelquist:1993ka} and a purely linear regime. In the former, non-renormalizable operators 
containing extra powers of a light $h$ are weighted by powers of $h/f$~\cite{Georgi:1984af}, while 
the Goldstone boson contributions encoded in $\UH(x)$ do not exhibit any scale suppression. In the linear regime, instead, the light $h$ and the three SM GBs are encoded into the scalar doublet $H$, with mass dimension one: therefore any extra insertion of $H$ is suppressed by a power of the cutoff.

It is becoming customary to parametrize the Lagrangian describing a light dynamical Higgs particle $h$ 
by means of the following ansatz~\cite{Contino:2010mh,Azatov:2012bz} (see also Ref.~\cite{Grinstein:2007iv}):
\begin{widetext}
\begin{align}
\LL_h=&\phantom{+\,\,}\frac{1}{2} (\partial_\mu h) (\partial^\mu h) \,\left(1+c_H\,\xi\,\cF_H(h)\right) 
   \,-\, V(h) \, + \, 
\nn \\
& - \frac{v^2}{4}\Tr\left[\VL^\mu \VL_\mu\right] \,\cF_C(h) \,+\, 
    c_T\,\xi\,\frac{v^2}{4}\, \Tr\left[\TL\VL^\mu\right]\Tr\left[\TL\VL_\mu\right] \cF_T(h) \, +\, \label{L3}\\
& - \left(\frac{v}{\sqrt{2}}\bar{Q}_L\UH(x)\,\cY\,\diag\left(\cF^U_Y(h),\cF^D_Y(h)\right) Q_R+\mbox{h.c.} \right) \,+ \, \ldots\,,\nn
\end{align}
\end{widetext}
where dots stand for higher order terms in the (linear) expansion in $h/f$, and \mbox{$\VL_\mu\equiv 
\left(\DL_\mu\UH\right)\UH^\dagger$} ($\TL\equiv\UH\sigma_3\UH^\dag$) is the vector (scalar) chiral field 
transforming in the adjoint of $SU(2)_L$. The covariant derivative reads 
\beq
\DL_\mu \UH(x) \equiv \derp_\mu \UH(x) +\dfrac{ig}{2}W_{\mu}^a(x)\sigma_a\UH(x) - 
                      \dfrac{ig'}{2} B_\mu(x) \UH(x)\sigma_3 \, , \nn
\eeq
with $W^a_\mu$ ($B_\mu$) denoting the $SU(2)_L$ ($U(1)_Y)$ gauge bosons and $g$ ($g'$) the corresponding 
gauge coupling. In the  equations above, $V(h)$ denotes the effective scalar potential describing the 
breaking of the electroweak symmetry. The first line in Eq.~(\ref{L3}) includes the Higgs kinetic term,  
while the second line describes the $W$ and $Z$ masses and their interactions with $h$, as well as the 
usual custodial symmetry breaking term labeled by $c_T$. Finally, restricting our considerations to the 
quark sector, the third line in Eq.~(\ref{L3}) accounts for the Yukawa-like interactions between $h$ and 
the SM quarks, grouped in doublets of the $SU(2)_{L,R}$ global symmetry $Q_{L,R}$, and with 
\beq
\cY\equiv\diag\left(Y_U,\,Y_D\right)\,,
\nn
\eeq
$Y_U$ and $Y_D$ being the usual Yukawa matrices.
The parameters $c_H$ and $c_T$ are model dependent operator coefficients.
 
The functions $\cF_H(h)$, $\cF_C(h)$, $\cF_T(h)$ and $\cF^{U,D}_Y(h)$ above, as well as all $\cF(h)$ functions to be 
used below, encode the generic dependence on $(\mean{h}+h)$ and are model-dependent. Each $\cF(h)$ function can 
be expanded in powers of $\xi$, $\cF(h)= g_0(h,v) + \xi g_1(h,v) + \xi^2 g_2(h,v) + \ldots$, where $g(h,v)$ 
are model-dependent functions of $h$ and of $v$, once $\mean{h}$ is expressed in terms of $\xi$ and $v$. For large $\xi$ the whole series may need to be considered.
In previous literature~\cite{Contino:2010mh,Azatov:2012bz} the functional dependence of some of those 
functions has been expressed as a power series in $h/v$: 
\beq
\begin{aligned}
\cF_C(h)&= \left(1+2a\,\frac{h}{v}+b\,\frac{h^2}{v^2} + \dots \right)\,,\\
\cF^{U,D}_Y(h)&=\left(1+c^{U,D}\,\dfrac{h}{v}+\ldots\right)\,.
\end{aligned}
\nn
\eeq
The constants $a$, $b$ and $c^{U,D}$ are model-dependent parameters and encode the dependence on $\xi$. 
The $a$ and $c_T$ parameters are constrained from electroweak precision tests: in particular 
$0.7\lesssim a\lesssim1.2$ \cite{Azatov:2012qz} and $-1.7\times10^{-3}<c_T\,\xi<1.9\times10^{-3}$ 
\cite{Giudice:2007fh} at $95\%$ CL.

The above Lagrangian can be very useful to describe an extended class of ``Higgs'' models, ranging from 
the SM  scenario with a linear Higgs  sector (for $\mean{h}=v$, $a=b=c^{U,D}=1$ and neglecting the higher order terms 
in $h$), to the technicolor-like ansatz (for $f\sim v$ and omitting all terms in $h$) and intermediate situations with a light scalar $h$ from composite/holographic Higgs models \cite{Dimopoulos:1981xc,Kaplan:1983fs,Kaplan:1983sm,Banks:1984gj,Georgi:1984ef,Georgi:1984af,Dugan:1984hq,Agashe:2004rs,Contino:2006qr,Gripaios:2009pe} (in general for $f\ne v$) up to dilaton-like scalar frameworks \cite{Halyo:1991pc,Goldberger:2007zk,Vecchi:2010gj,Campbell:2011iw,Matsuzaki:2012mk,Chacko:2012vm,Bellazzini:2012vz} (for $f\sim v$), where the dilaton participates to the electroweak symmetry breaking. Note that in concrete models electroweak corrections imply $\xi\lesssim 0.2-0.4$~\cite{Contino:2010rs}, but we will leave the $\xi$ parameter free here and account for the constraints on custodial symmetry through limits on the $d=2$ and higher-dimensional chiral operator coefficients.

In this work we analyze the strong interacting scenario in the presence of a light Higgs particle and construct 
the tower of pure gauge and gauge-$h$ operators up to four derivatives, in the context of the effective chiral Lagrangian. We will not consider neither pure Higgs operators, except for the $h$ kinetic term and the scalar potential, nor fermions operators that cannot be written in terms of pure-gauge or gauge-Higgs ones via the equations of motion, but for the standard Yukawa-type couplings.
We will assume a light $h$ and a  strong dynamics for the pseudo-Goldstone bosons which are the longitudinal degrees of freedom of the electroweak gauge bosons. This analysis enlarges and completes the operator basis previously considered in Refs.~\cite{Appelquist:1980vg,Longhitano:1980iz,Longhitano:1980tm,Feruglio:1992wf,Appelquist:1993ka,Contino:2010mh,Azatov:2012bz}.

%
%%%%%%%%%%%%%%%%%%%%%%%%%   2.  The effective Lagrangian   %%%%%%%%%%%%
%

\section{The effective Lagrangian}
\label{EffectiveLagrangian}

The parameter $\xi$, defined in Eq.~(\ref{xi}), encodes the strength of the effects at the electroweak scale for  
theories which exhibit strong coupling at the new physics scale $\Lambda_s\le 4\pi f$. Therefore, with a slight 
abuse of language $\xi$ measures the degree of non-linearity of the low-energy effective theory: $\xi 
\rightarrow 0$ refers to the linear regime, and $\xi \rightarrow 1 $ to the non-linear one. \\

\noindent{\bf Linear regime} \\

For $\xi \ll 1$ the hierarchy between $d\ge4$ effective operators mimics the linear expansion, where the 
operators are written in terms of the Higgs doublets H: couplings with higher number of (physical) Higgs 
legs are suppressed compared to the SM renormalizable ones, due to higher powers of $1/f$ or, in other 
words, of $\xi$. The power of $\xi$ keeps then track of the $h$-dependence of the higher dimension operators.

In the extreme linear limit $\mean{h} = v$, and the Higgs sector enters the tower of operators through powers of 
the SM Higgs doublet $H$ and its derivatives. It is illustrative to write  $H$ and its covariant derivative  
in terms of the Goldstone bosons matrix $\UH$ (where from now on the variable $x$ is left implicit) and the 
physical scalar $h$:
\beq
\begin{aligned}
H&=\dfrac{(v+h)}{\sqrt2}\,\UH\,\begin{pmatrix}0 \\ 1\end{pmatrix}\,,\\
\DL_\mu H&=\dfrac{(v+h)}{\sqrt2}\,\DL_\mu\UH\begin{pmatrix}0 \\ 1\end{pmatrix}+
\dfrac{\derp_\mu h}{\sqrt2}\,\UH\begin{pmatrix}0 \\ 1\end{pmatrix}\,,
\end{aligned}
\label{HiggDoubletDef}
\eeq
with $\DL_\mu\UH$ being the covariant derivative previously defined. The Higgs kinetic energy term in the 
linear expansion reads then:
\beq
(\DL^\mu H)^\dagger (\DL_\mu H) = \frac{1}{2}(\partial_\mu h)^2 - \frac{v^2}{4}\left(1+\frac{h}{v}\right)^2 
    \Tr\left[\VL_\mu\VL^\mu\right] \, . \nn
\eeq
On the right-hand side of this equation one can recognize the phenomenological Lagrangian in Eq.~(\ref{L3}) 
for  $f\rightarrow \infty$, i.e. $\xi=0$,  and $a=b=c=1$ (disregarding higher order terms in $h/f$), which 
corresponds to the SM case. A $(v+h)$ structure is clearly identified in the non-derivative term: the tower 
of $d>4$ operators would  inherit generically an $h$ dependence in powers of $(v+h)/f= \xi^{1/2} (1+h/v)$, and of 
$\partial_\mu h/f^2$~\cite{Giudice:2007fh,Contino:2010mh,Azatov:2012bz}. A priori, the $\cF(h)$ functions would 
also inherit that universal behavior in powers of $(1+h/v)$: for any operator weighted by $\xi^{n}$ it could 
be expected a dependence $\cF(h) = (1+h/v)^{2n}$. Nevertheless, the use of the equations of motion and 
integration by parts to construct the basis below will translate into combinations of operator coefficients, 
which lead to a generic $h$ dependence that, for instance at order $\xi$ (i.e for $d=6$ operators), reads
\beq
\cF_i(h)=\left(1+2\,a_i\,\frac{h}{v} + b_i\, \frac{h^2}{v^2}\right)\,,
\nn
\eeq
where $a_i$ and $b_i$ are expected to be $\cO(1)$. An obvious extrapolation applies to couplings weighted 
by higher powers of $\xi$ (i.e. for $d>6$ operators). In the following, all the discussion will be carried out 
in terms of generic $\cF(h)$ functions, though.\\

\noindent{\bf Non-linear regime} \\

For $\xi\approx1$, the $\xi$ dependence does not entail a suppression of operators compared to the renormalisable 
SM operators and the chiral expansion should instead be adopted, although it should be clarified at which level 
the effective expansion on $h/f$ should stop. In fact, for any BSM theory in the non-linear regime the dependence 
on $h$ will be a general function. For instance, in  the $SO(5)/SO(4)$ strong-interacting model with a composite 
light Higgs~\cite{Agashe:2004rs}, the tower of higher-dimension operators is weighted by powers of 
$\sin\left(\left(\mean{h}+h\right)/f\right)$, and in this case $\xi= \sin^2\left(\mean{h}/f\right)$.
Below, the $\cF(h)$ functions encode the non-linear interactions of the light $h$ and will be considered completely general polynomials of $\mean{h}$ and $h$ (not including derivatives of $h$). Notice that, when using the equations of motion and integration by parts to relate operators, $\cF(h)$ would be assumed to be redefined when convenient, much as one customarily redefines the constant operator coefficients.

%%%%%%%%%%%%%%%%%%%%%%%%%   2.1  Basis of operators in the non-linear regime
\boldmath
\subsection{Pure gauge and gauge-$h$ operator basis}
\unboldmath

All CP-even operators appropriate to the non-linear regime will be included in this work, up to four derivatives. 

The connection to the linear regime will be made manifest exploiting the operator dependence on $\xi$. The Lagrangian can be decomposed as 
\beq
\LL_{gauge-h}= \LL_{\chi=0}^h + \LL_{\chi=2}^h +  \LL_{\chi=3}^h + \LL_{\chi=4}^h \,,
\label{Ltotal}
\nn
\eeq
where the subscript $\chi=n$ reminds the dimension of the non-linear parenthood of the operators. 
$\LL^h_{\chi=0}$ contains kinetic energy-type Higgs couplings plus the scalar potential and corresponds to the first line in 
Eq.~(\ref{L3}). $\LL_{\chi=2}^h$ accounts for the $W$ and $Z$ boson masses and their interactions 
with the $h$ field, and is given in the second line in Eq.~(\ref{L3}). $\LL_{\chi=3}^h$ 
is the Yukawa-type coupling and corresponds to the third line in Eq.~(\ref{L3}).
Finally, the $\LL_{\chi=4}^h$ term can be written as: 
\begin{widetext}
\beq
\begin{split}
\LL_{\chi=4}^h=&
-\frac{1}{4}\,G_{\mu\nu}^aG^{\mu\nu}_a\,\left(1+c_G\,\xi\,\cF_G(h)\right)
-\frac{1}{4}\,W_{\mu\nu}^aW^{\mu\nu}_a\,\left(1+c_W\,\xi\,\cF_W(h)\right)
-\frac{1}{4}\,B_{\mu\nu}B^{\mu\nu}\,\left(1+c_B\,\xi\,\cF_B(h)\right)+\\
&+\xi\, \sum_{i=1}^{5} \,c_i\,\cP_i(h)\,
+\, \xi^2 \,\sum_{i=6}^{22}\, c_i\,\cP_i(h)
+\, \xi^3 \,\sum_{i=23}^{25}\, c_i\,\cP_i(h)
+\, \xi^4 \, c_{26}\,\cP_{26}(h) \,,
\end{split}
\label{L4}
\eeq
\end{widetext}
The first line in Eq.~(\ref{L4}) includes the kinetic terms for the gauge bosons, with $W_{\mu\nu}$, 
$B_{\mu\nu}$ and $G_{\mu\nu}$ denoting the $SU(2)_L$, $U(1)_Y$ and $SU(3)_C$ field strengths, respectively. 
The second line in Eq.~(\ref{L4}) contains the following $26$ CP-even operators, ordered by their $\xi$ dependence:
\bea
&\hspace{-2.1cm}
\begin{aligned}
\blue{\cP_{1}(h)}\,\,  &= g\,g' \,B_{\mu\nu} \Tr\left(\TL\,W^{\mu\nu}\right)\,\cF_{1}(h)\\ 
\blue{\cP_{2}(h)}\,\,  &= i\,g' \,B_{\mu\nu} \Tr\left(\TL\left[\VL^\mu,\VL^\nu\right]\right)\,\cF_{2}(h) \\
\blue{\cP_{3}(h)}\,\,  &= i\,g\,\Tr\left(W_{\mu\nu}\left[\VL^\mu,\VL^\nu\right]\right)\,\cF_{3}(h) \\
\red{\cP_{4}(h)}\,\,  &= i\,g' B_{\mu\nu}\Tr(\TL\VL^\mu)\,\derp^\nu \cF_{4}(h) \\
\red{\cP_{5}(h)}\,\,  &= i\,g \,\Tr(W_{\mu\nu}\VL^\mu)\,\derp^\nu \cF_{5}(h) \\
\end{aligned}
\label{GaugeOperators1d5}
\\
\nn\\
&\hspace{-0.7cm}
\begin{aligned}
\blue{\cP_{6}(h)}\,\, &=\left(\Tr\left(\VL_\mu\,\VL^\mu\right)\right)^2\,\cF_{6}(h) \\
\blue{\cP_{7}(h)}\,\, &=\left(\Tr\left(\VL_\mu\,\VL_\nu\right)\right)^2\,\cF_{7}(h) \\
\blue{\cP_{8}(h)}\,\, &=g^2\,\left(\Tr\left(\TL\,W^{\mu\nu}\right)\right)^2\,\cF_{8}(h) \\
\blue{\cP_{9}(h)}\,\, &=i\,g\,\Tr\left(\TL\,W_{\mu\nu}\right)\Tr\left(\TL\left[\VL^\mu,\VL^\nu\right]\right)\,
                       \cF_{9}(h)\\  
\blue{\cP_{10}(h)} &=g\,\epsilon^{\mu\nu\rho\lambda}\Tr\left(\TL\VL_\mu\right)\Tr\left(\VL_\nu\,W_{\rho\lambda}
                       \right)\,\cF_{10}(h)\\
\green{\cP_{11}(h)} &= \Tr\left((\cD_\mu\VL^\mu)^2 \right)\,\cF_{11}(h) \\
\green{\cP_{12}(h)} &= \Tr(\TL\,\cD_\mu\VL^\mu)\,\Tr(\TL\,\cD_\nu\VL^\nu)\,\cF_{12}(h) \\
\green{\cP_{13}(h)} & = \Tr([\TL \,,\VL_\nu]\,\cD_\mu \VL^\mu) \, \Tr(\TL\VL^\nu) \cF_{13}(h) \\
\red{\cP_{14}(h)} &= i\,g \,\Tr(\TL W_{\mu\nu})\Tr(\TL\VL^\mu)\,\derp^\nu \cF_{14}(h)\\
\red{\cP_{15}(h)} &=\Tr(\TL\,[\VL_\mu,\VL_\nu])\Tr(\TL\VL^\mu)\,\derp^\nu \cF_{15}(h)\\
\red{\cP_{16}(h)} &=\Tr(\VL_\nu \,\cD_\mu\VL^\mu)\,\derp^\nu \cF_{16}(h)\\
\red{\cP_{17}(h)} &=\Tr(\TL\,\cD_\mu\VL^\mu)\Tr(\TL\VL_\nu)\,\derp^\nu \cF_{17}(h) \\
\red{\cP_{18}(h)} &=\Tr\left(\VL_\mu\,\VL^\mu\right)\,\derp_\nu\derp^\nu\cF_{18}(h)\\
\red{\cP_{19}(h)} &=\Tr(\VL_\mu \VL^\mu)\,\derp_\nu\cF_{19}(h)\derp^\nu\tilde\cF_{19}(h)\\
\red{\cP_{20}(h)} &=\Tr\left(\VL_\mu\,\VL_\nu\right)\,\derp^\mu\cF_{20}(h)\derp^\nu\tilde\cF_{20}(h)\\
\red{\cP_{21}(h)} &=\left(\Tr\left(\TL\VL_\mu\right)\right)^2 \derp_\nu\cF_{21}(h)\derp^\nu\tilde\cF_{21}(h)\\
\red{\cP_{22}(h)} &=\Tr\left(\TL\VL_\mu\right)\Tr\left(\TL\VL_\nu\right)\,\derp^\mu\cF_{22}(h)\derp^\nu\tilde\cF_{22}(h)
\end{aligned} 
\label{GaugeOperators6d22} \\ 
\nn \\
&\hspace{-0.8cm}
\begin{aligned}
\blue{\cP_{23}(h)} &= \Tr\left(\VL_\mu\VL^\mu\right)\left(\Tr\left(\TL\VL_\nu\right)\right)^2\cF_{23}(h) \\
\blue{\cP_{24}(h)} &=\Tr\left(\VL_\mu\VL_\nu\right)\Tr\left(\TL\VL^\mu\right)\Tr\left(\TL\VL^\nu\right)\cF_{24}(h)\\
\red{\cP_{25}(h)} &=\left(\Tr\left(\TL\,\VL_\mu\right) \right)^2 \derp_\nu\derp^\nu\cF_{25}(h)
\end{aligned}
\label{GaugeOperators23d25}\\
\nonumber\\
&\hspace{-1.5cm}
\begin{aligned}
\blue{\cP_{26}(h)} &=\left(\Tr\left(\TL\VL_\mu\right)\Tr\left(\TL\VL_\nu\right)\right)^2\,\cF_{26}(h)\,.
\end{aligned}
\label{GaugeOperators26}
\eea
The $26$ constant parameters $c_i$ are model-dependent coefficients. The powers of $\xi$, factorized out in 
the second line of Eq.~(\ref{L4}), do {\bf not} reflect an expansion in $\xi$, but a reparametrisation that facilitates the tracking to the lowest dimension at which a ``sibling'' operator appears in the linear expansion. By sibling we mean an operator written in terms of the Higgs 
doublet $H$, that includes the pure gauge part of the couplings $\cP_{1-26}(h)$. It may happen that an 
operator listed in Eqs.~(\ref{GaugeOperators1d5})-(\ref{GaugeOperators26}) corresponds to a specific 
combination of siblings with different dimensions. This is the case, for instance, of $\cP_{13}(h)$, 
whose linear siblings are of dimension $8$ and $10$.

For $\xi\ll 1$ the weight of the operators which are accompanied by powers of $\xi$ is scale suppressed 
compared to that of SM renormalisable couplings. In this limit the Lagrangian above would, for particular combination of the coefficients $c_i$, correspond to the linear expansion up to $d=6$ operators, if only the terms of zero and first order in $\xi$ are kept. Operators  $\cP_{6}(h)$ to $\cP_{26}(h)$ would correspond instead to $d=8$ or higher-dimension siblings in the linear expansion. In contrast, in the non-linear regime, that is for $\xi\approx 1$, no such suppression appears 
and {\it all} operators in Eqs.~(\ref{GaugeOperators1d5})-(\ref{GaugeOperators26}) should be considered on equal footing. The leading terms of the linear and non-linear 
expansions do not match.

Operators in Eq.~(\ref{L3}) and in the first line of Eq.~(\ref{L4}), as well as $\cP_{1-5}(h)$ had been already 
pointed out in the analysis of the linear-non linear connection of the SILH framework~\cite{Giudice:2007fh}. Indeed, in the limit of small $\xi$, we can safely neglect all the terms proportional to $\geq2$ powers of $\xi$ and the resulting Lagrangian has a correspondence with the SILH one.
Nevertheless, to be complete the rest of the operators mentioned above should be included when fermions are taken into account and/or when dealing with theories in the non-linear regime. Equivalently in the linear regime, one should consider operators with $d>6$: the complete basis of operators in this case accounts for operators of $d=12$ at most, while all the higher order operators are redundant. This is consistent with the basis in the non-linear regime presented here, where the lowest dimensional sibling of $\cP_{26}(h)$ has indeed dimension 12.

The different operators defined in Eqs.~(\ref{GaugeOperators1d5})-(\ref{GaugeOperators26}) correspond 
to three major categories: pure gauge and gauge-$h$ operators (in blue) which result from a direct extension of the original Appelquist-Longhitano chiral Higgsless  basis; operators containing the contraction 
$\cD_\mu \VL^\mu$ and no derivatives of $\cF(h)$ (in green); operators with one or two derivatives of 
$\cF(h)$ (in red).\\

\noindent{\bf \boldmath The extended Appelquist-Longhitano basis}\\

$\cP_{1-3}(h)$, $\cP_{6-10}(h)$, $\cP_{23-24}(h)$ and $\cP_{26}(h)$ result from combining the basis of independent $d=4$ chiral operators already considered in Refs.~\cite{Appelquist:1980vg,Longhitano:1980iz,Longhitano:1980tm,
Feruglio:1992wf,Appelquist:1993ka} with additional $\cF(h)$ insertions. They appear in the Lagrangian with different 
powers of $\xi$: $\cP_{1-3}(h)$ is linear in $\xi$, while $\cP_{6-10}(h)$, $\cP_{23-24}(h)$ 
and $\cP_{26}(h)$ are proportional to $\xi^2$, $\xi^3$ and $\xi^4$, respectively. 

This ensemble constitutes a complete basis of linearly independent pure gauge and gauge-$h$ operators with at most four derivatives, when one does not consider neither derivatives of $h$ nor fermion masses or fermionic operators that cannot be related to pure-gauge or gauge-Higgs ones via the equations of motion. It is worth noticing that, neglecting all terms in $h$ (i.e. taking $\cF(h)$ as a constant), the list of operators in Eqs.~(\ref{GaugeOperators1d5})-(\ref{GaugeOperators26}) reduces to the original Appelquist-Longhitano basis.\\

\noindent{\bf Derivatives of \boldmath $h$}\\

Terms resulting from combining $\derp_\mu h$ or $\derp_\mu\derp^\mu h$ with  $d=2$ or $d=4$  chiral couplings 
enlarge the basis in the previous paragraph by several operators: $\cP_{4-5}(h)$, $\cP_{14-22}(h)$ and $\cP_{25}(h)$. 

Among all the operators of this class, three of them have been already identified in Ref.~\cite{Giudice:2007fh,Contino:2010mh,Azatov:2012bz}. We have provided the full set of $12$ operators with derivatives of $h$ that need to be taken into account to complete the pure gauge and gauge-$h$ basis; they exhibit a $\xi$-dependence which starts at the linear, quadratic or cubic level. \\

\noindent{\bf\boldmath Massive fermions}\\

Several operators in the list in Eqs.~(\ref{GaugeOperators1d5})-(\ref{GaugeOperators26}) are physical only in the presence of massive fermions: these are $\cP_{11-13}(h)$, $\cP_{16}(h)$, $\cP_{17}(h)$, $\cP_{19}(h)$ and $\cP_{21}(h)$. In particular, $\cP_{11-13}(h)$, $\cP_{16}(h)$ and $\cP_{17}(h)$ contain the contraction $\cD_\mu\,V^\mu$ that can be shown to be connected to the Yukawa couplings. Indeed, considering the equations of motion for the field strengths\footnote{We thank Concha Gonzalez-Garcia for having asked the differences among the equations of motion in the linear and in the non-linear regime. This led us to correct typos on the equations of motion and expressions connected to them.},
\beq
\begin{aligned}
\left(D^\mu\,W_{\mu\nu}\right)_j &= i\dfrac{g}{4}v^2\Tr[\VL_\nu\,\sigma_j] \cF_C(h)+
\dfrac{g}{2}\bar Q_L\gamma_\nu\sigma_jQ_L \,,\\
\derp^\mu B_{\mu\nu}&= -i\dfrac{g'}{4}v^2\Tr[\TL\,\VL_\nu] \cF_C(h)+ 
g'\sum_{i=L,R} \bar Q_i\gamma_\nu {\mathbf h}_i Q_i\,, 
\end{aligned}
\nn
\eeq
with ${\mathbf h}_{L,R}$ the left and right hypercharges in the $2\times 2$ matrix notation, and deriving 
these expressions, a connection is established between operators containing $\cD_\mu\VL^\mu$ and fermionic 
currents that preserve flavour but change chirality:
\beq
\begin{aligned}
\frac{i v}{\sqrt{2}}\Tr(\sigma_j \,\cD_\mu\VL^\mu) \cF_C(h)= &
-\frac{i v}{\sqrt{2}}\Tr(\sigma_j \VL^\mu)\derp_\mu\cF_C(h)+\\
&\hspace{-2.4cm}+ i\,{\bar Q}_L\,\sigma_j\,\UH\,\cY\diag\left(\cF^U_Y(h),\,\cF^D_Y(h)\right)Q_R + {\rm h.c.}\,,\\
\frac{i v}{\sqrt{2}}\Tr(\TL\,\cD_\mu\VL^\mu)\cF_C(h) = &
-\frac{i v}{\sqrt{2}}\Tr(\TL\,\VL^\mu)\derp_\mu\cF_C(h)+\\
&\hspace{-2.4cm}+i\,{\bar Q}_L\,\TL\,\UH\, \cY\diag\left(\cF^U_Y(h),\,\cF^D_Y(h)\right)Q_R + {\rm h.c.}\,,
\end{aligned} \nn
\eeq
where the relation $\cD_\mu \TL = [\VL_\mu , \TL]$ and the Dirac equations
\beq
\begin{aligned}
&i \slashed{D}_L Q_L = \frac{v}{\sqrt{2}}\,\UH \,\cY\diag\left(\cF^U_Y(h),\cF^D_Y(h)\right) Q_R \,,\\
&i \slashed{D}_R Q_R =  \frac{v}{\sqrt{2}}\,\cY^\dag\diag\left(\cF^U_Y(h),\,\cF^D_Y(h)\right) \UH^\dag\, Q_L\,,  
\end{aligned} \nn
\eeq
with $\slashed{D}_{L,R}$ the usual covariant derivatives acting on the $L,R$ doublet spinors, have been used. As a result, if fermion masses are neglected, $\cP_{11,16}(h)$, $\cP_{12,17}(h)$ and $\cP_{13}(h)$ can be rewritten in terms of $\cP_{20}(h)$, $\cP_{22}(h)$ and $\cP_{15}(h)$, respectively.

Furthermore, using the equation of motion for the light $h$,
\beq
\begin{split}
-\derp_\mu\derp^\mu h &= \dfrac{v^2}{4}\Tr\left[V_\mu V^\mu\right] \dfrac{\derp \cF_C(h)}{\derp h} +\dfrac{\derp V(h)}{\derp h}+\\
&+\dfrac{v}{\sqrt2}\left({\bar Q}_L\,\UH\,\cY\diag\left(\cF^U_Y(h),\,\cF^D_Y(h)\right)Q_R + {\rm h.c.}\right)\,,
\end{split}\nn
\eeq
operator $\cP_{19}$ ($\cP_{21}$) can be reduced to a combination of $\cP_{6}(h)+\cP_{18}(h)$ ($\cP_{23}(h)+\cP_{25}(h)$), plus a term that can be absorbed in the redefinition of the couplings of the gauge bosons and $h$, plus a term containing the Yukawa interactions.

In consequence, if fermion masses are neglected, $\cP_{11-13}(h)$, $\cP_{16}(h)$,  $\cP_{17}(h)$, $\cP_{19}(h)$ and $\cP_{21}(h)$ become redundant. Conversely, these operators are independent from the other operators of the basis and should be taken into account. These operators together with the pure gauge and gauge-$h$ ones in the classes previously defined, constitute a complete basis of linearly independent operators with at most four derivatives, upon disregarding: i) flavour-changing fermionic operators \cite{Alonso:2012jc,Alonso:2012pz} but the Yukawa coupling; ii) flavour-conserving fermionic operators that cannot be related to pure-gauge or gauge-Higgs ones via the equations of motion; iii) pure-$h$ effective couplings. \\

\noindent{\bf Custodial symmetry nature}\\

In the list in Eqs.~(\ref{GaugeOperators1d5})-(\ref{GaugeOperators26}), the operators $\cP_{1}$, $\cP_{2}$, $\cP_{4}$, $\cP_{8-15}$, $\cP_{17}$, and $\cP_{21-26}$ are custodial symmetry breaking. This can be understood either by the presence of the hypercharge coupling constant $g'$ in front of the operators $\cP_{1}$, $\cP_{2}$, and $\cP_{4}$, or by the connection to quark masses, as it is the case for $\cP_{11-13}$, $\cP_{16}$ and $\cP_{17}$, or finally through the presence of the chiral scalar field $\TL$ that explicitly violates the custodial symmetry.

\subsection{Connection with other bases}

The list in Eqs.~(\ref{GaugeOperators1d5})-(\ref{GaugeOperators26}) accounts for pure gauge and gauge-$h$ operators with at most four derivatives. Any coupling with a light $h$ is weighted by the appropriate powers of the scale $f$ and encoded in the generic adimensional $\cF(h)$ functions. If instead a counting based on the canonical dimension is performed, the lowest dimensional couplings which include $h$ have canonical dimension $5$ for all the operators, except for $P_{19-22}$ that have $d=6$.

It is easy to establish the correlation between the basis defined above and other possible gauge or gauge-$h$ 
bases of operators with at most four derivatives, present in the literature.
To this aim, two equalities are useful:
\beq
\begin{gathered}
\VL_{\mu\nu} \equiv \cD_{\mu}\VL_{\nu}-\cD_{\nu}\VL_{\mu} = 
ig\WL_{\mu \nu} 
-i\frac{g'}{2}B_{\mu\nu}\TL+ \left[\VL_\mu, \VL_\nu\right] \,, \\
\left[\cD_\mu,\,\cD_\nu\right]\cO = ig[W_{\mu\nu}, \cO]\,,
\end{gathered}
\nonumber
\eeq
where $\cO$ is a generic operator covariant under $SU(2)_L$ and invariant under $U(1)_Y$.

Two operators with two derivatives acting on the generic functions $\cF_i(h)$ can be written, in addition 
to $\cP_{18-22}(h)$ and $\cP_{25}(h)$. However, via integration by parts and pertinent redefinition of the generic 
functions $\cF_i(h)$, one obtains that two of these new structures are given by:
\begin{widetext}
\begin{align}
&\Tr\left(\VL_\mu\,\VL_\nu\right)\derp^\mu\derp^\nu\cF(h)=\frac{1}{2}\cP_{4}(h)-\cP_{5}(h)-\cP_{16}(h) 
       + \dfrac{1}{2}\cP_{18}(h)\,,
       \label{Contino1}\\
&\Tr\left(\TL\VL_\mu\right)\Tr\left(\TL\,\VL_\nu\right)\derp^\mu\derp^\nu\cF(h)=\cP_{4}(h)-\cP_{14}(h)+\cP_{15}(h)-\cP_{17}(h)
       +\frac{1}{2}\cP_{25}(h)\,.
\end{align}
\end{widetext}
%The remaining two, $\left(\Tr\left(\TL\VL_\mu\right)\right)^2 \derp_\nu\cF(h)\derp^\nu\cF'(h)$ and $\Tr(\VL_\mu \VL^\mu)\derp_\nu\cF(h)\derp^\nu\cF'(h)$, can be reduced to $\cP_{25}(h)$ and $\cP_{18}(h)$, respectively, by the use of the $h$ equation of motion. The latter operator and the one in Eq.~(\ref{Contino1}) have been introduced in Ref~\cite{Azatov:2012bz}.

Next, operators containing derivatives of the field strengths can be decomposed as  
\begin{widetext}
\beq
\begin{aligned}
&ig'\left(\partial_\mu B^{\mu\nu}\right)\Tr \left[\TL\,\VL_{\!\nu}\right]\,\cF(h) = 
  -\dfrac{g^{\prime2}}{2}\,B_{\mu\nu}B^{\mu\nu}\cF(h)+\dfrac{1}{2}\,\cP_{1}(h)+\frac{1}{2}\,\cP_{2}(h)+\cP_{4}(h)
\,,\\
&ig\,\Tr \left[\left(\cD_\mu W^{\mu\nu}\right)\VL_{\!\nu} \right] \,\cF(h)= 
\dfrac{g^2}{4}\,W^{a}_{\mu\nu}W_{a}^{\mu\nu}\cF(h)-\dfrac{1}{4}\,\cP_{1}(h)-\frac{1}{2}\,\cP_{3}(h)+\cP_{5}(h)
\,,\nn
\end{aligned}
\eeq
\beq
ig\,\Tr\left[\left(\cD_\mu W^{\mu\nu}\right)\TL \right]\Tr\left[\TL \,\VL_{\!\nu} \right]\cF(h)=
-\dfrac{1}{2}\cP_{1}(h)-\cP_{3}(h)+\dfrac{1}{2}\cP_{8}(h)+\cP_{9}(h)+\cP_{14}(h)\,.
\nn
\eeq
\end{widetext}

Finally, disregarding the dependence on the $h$ field, the three operators $\cP_{11-13}$ containing the 
contraction $\cD_\mu \VL^\mu$ have already been considered in Ref.~\cite{Feruglio:1992wf}, although with 
a slightly different notation for the last two. The relation among $\cP_{12}$ and $\cP_{13}$ and the 
corresponding operators in Ref.~\cite{Feruglio:1992wf} is the following:
\begin{widetext}
\beq
\begin{aligned}
&\Tr\left(\TL\,\cD_\mu\,\cD_\nu\,\VL^\nu\right)\,\Tr\left(\TL\,\VL^\mu\right)\,\cF(h)=
-\cP_{12}(h)+\cP_{13}(h)-\cP_{17}(h)\,, \\
&\begin{split}
\Tr\left(\TL\,\cD_\mu\,\VL_\nu\right)\Tr\left(\TL\,\cD^\mu\,\VL^\nu\right)\cF(h)=&
-\frac{g^{\prime2}}{2}B_{\mu\nu}B^{\mu\nu}\,\cF(h)+\cP_{1}(h)+\cP_{4}(h) -2\cP_{6}(h)+2\cP_{7}(h)-
    \frac{1}{2}\cP_{8}(h)+\\
&+\cP_{12}(h) -\cP_{14}(h)+\cP_{15}(h)+\cP_{17}(h) +2\cP_{23}(h)-2\cP_{24}(h) +\frac{1}{2}\cP_{25}(h)\,.
\end{split}
\end{aligned}
\nonumber
\eeq
\end{widetext}

%
%%%%%%%%%%%%%%%%%%%%%%%%%   4.  Conclusions       %%%%%%%%%%%%%%%%%%%%%%%%
%

\section{Conclusions}
\label{Conclusions}

In this paper, we have considered the generic scenario in which a strong dynamics lies behind a light Higgs particle $h$, 
within an effective Lagrangian approach. The parameter describing the degree of non-linearity $\xi=(v/f)^2$ must lie 
in the range $0<\xi<1$. Small values lead to a low-energy theory undistinguishable from the SM, since all the effects 
of the strong interacting theory at the high scale become negligible. Larger values indicate a chiral regime for the 
dynamics of the Goldstone bosons, which in turn requires to use a chiral expansion to describe them, combined with 
appropriate insertions of the light $h$ field. 

This work generalizes the operator basis of Refs.~\cite{Appelquist:1980vg,Longhitano:1980iz,Longhitano:1980tm,
Feruglio:1992wf,Appelquist:1993ka} of chiral pure-gauge operators to include a light strong interacting $h$ 
particle, up to four derivatives operators for pure gauge and gauge-$h$ couplings. The complete basis obtained includes several supplementary operators with 
respect to those previously identified in the literature \cite{Giudice:2007fh,Low:2009di,Contino:2010mh,
Azatov:2012bz}, which need to be taken into account when approaching the non-linear regime. Furthermore, 
the results have been presented making explicit the leading dependence on $\xi$ for each operator, which 
allows a direct identification of the equivalent leading operator of the linear regime. The consideration 
of $d=6$, $8$, $10$ and $12$ couplings of the linear expansion turns out to be required to 
establish the connection with the set of operators of the non-linear one considered here.

These results may also provide a model-independent guideline to explore which exotic gauge-Higgs couplings 
may be expected, and their relative strength to Higgsless observable amplitudes. Complementary information 
could come from the flavour sector \cite{Alonso:2012jc,Alonso:2012pz} and hopefully will be able 
to shed light on the origin of the electroweak symmetry breaking mechanism.\\

%
%%%%%%%%%%%%%%%%%%%%%%%%%    Note added in proof       %%%%%%%%%%%%%%%%%%%%%%
%

\section*{Note added in proof}

After version 2 of the present manuscript appeared on the web, Ref.~\cite{Buchalla:2013rka} extended our work proposing a complete basis of all possible independent operators of the non-linear Lagrangian with a light Higgs particle, in a chiral expansion up to four derivatives; a particular selection of both bosonic and fermionic operators was chosen there. It also included the derivation of the corresponding EOMs for their specific choice of LO Lagrangian, which excluded some two-derivative operators for phenomenological reasons.

Ref.~\cite{Buchalla:2013rka} also contained some inferred criticisms to the results presented here, pointing to some allegedly missing and redundant operators. This criticism is incorrect: in the present manuscript the focus is set by definition on the pure gauge and gauge-$h$ couplings, that is, on defining a maximal set of such independent (and thus complete and non-redundant set of) operators: a basis for pure gauge and gauge-$h$ couplings. Those operators criticised by Ref.~\cite{Buchalla:2013rka} as ``missing" are not in this category; a similar comment applies to the redundancy issue, explained by the choice in Ref.~\cite{Buchalla:2013rka} of trading some gauge and gauge-$h$ operators by fermionic ones. Finally, the $\xi$ weights and the truncations defined here for the first time lead to rules for operator weights consistent with those defined long ago in the Georgi-Manohar counting~\cite{Manohar:1983md}, and more recently in Ref.~\cite{Jenkins:2013sda}.

%
%
%The inferred criticisms in Ref.~\cite{Buchalla:2013rka} to the results presented here about missing and redundant operators are incorrect: here, we concentrated by definition in pure gauge and gauge-$h$ couplings and those criticised as ``missing" are not in this category; a similar comment applies to the redundancy issue, explained by the choice of trading some gauge operators by fermionic ones in Ref.~\cite{Buchalla:2013rka}. Finally, the $\xi$ weights and the truncations defined here for the first time lead to rules for operator weights consistent with those defined long ago in the Georgi-Manohar counting~\cite{Manohar:1983md}, and more recently in Ref.~\cite{Jenkins:2013sda}.
%

%%%%%%%%%%%%%%%%%%%%%%%%%%%%%%%%%%%%%%%%%%%%%%%%%%%%%%%%%%%%
% Acknowledgements
%%%%%%%%%%%%%%%%%%%%%%%%%%%%%%%%%%%%%%%%%%%%%%%%%%%%%%%%%%%%
\section*{Acknowledgements}
We thank C.~Delaunay, C.~Grojean, J.~Perez and W.~Skiba for interesting questions and comments on the first version of the paper. We acknowledge partial support by European Union FP7 ITN INVISIBLES (Marie Curie Actions, PITN-GA-2011-289442), 
CiCYT through the project FPA2009-09017, CAM through the project HEPHACOS P-ESP-00346, European Union FP7 
ITN UNILHC (Marie Curie Actions, PITN-GA-2009-237920), MICINN through the grant BES-2010-037869 and the Juan 
de la Cierva programme (JCI-2011-09244), Italian Ministero 
dell'Uni\-ver\-si\-t\`a e della Ricerca Scientifica through the COFIN program (PRIN 2008) and the contracts 
MRTN-CT-2006-035505 and  PITN-GA-2009-237920 (UNILHC). R.A. and L.M. thank the Galileo Galilei Institute for Theoretical 
Physics for the hospitality and the INFN for partial support during the completion of this work. R.A. acknowledges the Harvard Physics department for hospitality during the completion phase of this work. S.R. and J.Y acknowledge CERN TH department for hospitality during the completion phase of the work. 

%%%%%%%%%%%%%%%%%%%%%%%%%%%%%%%%%%%%%%%%%%%%%%%%%%%%%%%%%%%%%%%%%
%%%%%%%%%%%%%%%%%%%%%%%%%  Bibliography     
%%%%%%%%%%%%%%%%%%%%%%%%%%%%%%%%%%%%%%%%%%%%%%%%%%%%%%%%%%%%%%%%%
%
%\bibliography{biblio}{}

\begin{thebibliography}{10}

\bibitem{:2012gk}
{\bf ATLAS} Collaboration, G.~Aad {\em et.~al.},  Phys.Lett. {\bf B716} (2012)
  1--29, [\href{http://xxx.lanl.gov/abs/1207.7214}{{\tt arXiv:1207.7214}}].

\bibitem{:2012gu}
{\bf CMS} Collaboration, S.~Chatrchyan {\em et.~al.},  Phys.Lett. {\bf B716}
  (2012) 30--61, [\href{http://xxx.lanl.gov/abs/1207.7235}{{\tt
  arXiv:1207.7235}}].

\bibitem{Englert:1964et}
F.~Englert and R.~Brout,  Phys.Rev.Lett. {\bf 13} (1964) 321--323.

\bibitem{Higgs:1964ia}
P.~W. Higgs,  Phys.Lett. {\bf 12} (1964) 132--133.

\bibitem{Higgs:1964pj}
P.~W. Higgs,  Phys.Rev.Lett. {\bf 13} (1964) 508--509.

\bibitem{Manohar:1983md}
A.~Manohar and H.~Georgi,  Nucl.Phys. {\bf B234} (1984) 189.

\bibitem{Susskind:1978ms}
L.~Susskind,  Phys. Rev. {\bf D20} (1979) 2619--2625.

\bibitem{Dimopoulos:1979es}
S.~Dimopoulos and L.~Susskind,  Nucl. Phys. {\bf B155} (1979) 237--252.

\bibitem{Dimopoulos:1981xc}
S.~Dimopoulos and J.~Preskill,  Nucl.Phys. {\bf B199} (1982) 206.

\bibitem{Kaplan:1983fs}
D.~B. Kaplan and H.~Georgi,  Phys.Lett. {\bf B136} (1984) 183.

\bibitem{Kaplan:1983sm}
D.~B. Kaplan, H.~Georgi, and S.~Dimopoulos,  Phys. Lett. {\bf B136} (1984) 187.

\bibitem{Banks:1984gj}
T.~Banks,  Nucl.Phys. {\bf B243} (1984) 125.

\bibitem{Georgi:1984ef}
H.~Georgi, D.~B. Kaplan, and P.~Galison,  Phys.Lett. {\bf B143} (1984) 152.

\bibitem{Georgi:1984af}
H.~Georgi and D.~B. Kaplan,  Phys.Lett. {\bf B145} (1984) 216.

\bibitem{Dugan:1984hq}
M.~J. Dugan, H.~Georgi, and D.~B. Kaplan,  Nucl. Phys. {\bf B254} (1985) 299.

\bibitem{Contino:2010rs}
R.~Contino,  \href{http://xxx.lanl.gov/abs/1005.4269}{{\tt arXiv:1005.4269}}.

\bibitem{Appelquist:1980vg}
T.~Appelquist and C.~W. Bernard,  Phys. Rev. {\bf D22} (1980) 200.

\bibitem{Longhitano:1980iz}
A.~C. Longhitano,  Phys. Rev. {\bf D22} (1980) 1166.

\bibitem{Longhitano:1980tm}
A.~C. Longhitano,  Nucl. Phys. {\bf B188} (1981) 118.

\bibitem{Feruglio:1992wf}
F.~Feruglio,  Int.J.Mod.Phys. {\bf A8} (1993) 4937--4972,
  [\href{http://xxx.lanl.gov/abs/hep-ph/9301281}{{\tt hep-ph/9301281}}].

\bibitem{Appelquist:1993ka}
T.~Appelquist and G.-H. Wu,  Phys.Rev. {\bf D48} (1993) 3235--3241,
  [\href{http://xxx.lanl.gov/abs/hep-ph/9304240}{{\tt hep-ph/9304240}}].

\bibitem{Contino:2010mh}
R.~Contino, C.~Grojean, M.~Moretti, F.~Piccinini, and R.~Rattazzi,  JHEP {\bf
  1005} (2010) 089, [\href{http://xxx.lanl.gov/abs/1002.1011}{{\tt
  arXiv:1002.1011}}].

\bibitem{Azatov:2012bz}
A.~Azatov, R.~Contino, and J.~Galloway,  JHEP {\bf 1204} (2012) 127,
  [\href{http://xxx.lanl.gov/abs/1202.3415}{{\tt arXiv:1202.3415}}].

\bibitem{Grinstein:2007iv}
B.~Grinstein and M.~Trott,  Phys.Rev. {\bf D76} (2007) 073002,
  [\href{http://xxx.lanl.gov/abs/0704.1505}{{\tt arXiv:0704.1505}}].

\bibitem{Azatov:2012qz}
A.~Azatov and J.~Galloway,  Int.J.Mod.Phys. {\bf A28} (2013) 1330004,
  [\href{http://xxx.lanl.gov/abs/1212.1380}{{\tt arXiv:1212.1380}}].

\bibitem{Giudice:2007fh}
G.~F. Giudice, C.~Grojean, A.~Pomarol, and R.~Rattazzi,  JHEP {\bf 06} (2007)
  045, [\href{http://xxx.lanl.gov/abs/hep-ph/0703164}{{\tt hep-ph/0703164}}].

\bibitem{Agashe:2004rs}
K.~Agashe, R.~Contino, and A.~Pomarol,  Nucl.Phys. {\bf B719} (2005) 165--187,
  [\href{http://xxx.lanl.gov/abs/hep-ph/0412089}{{\tt hep-ph/0412089}}].

\bibitem{Contino:2006qr}
R.~Contino, L.~Da~Rold, and A.~Pomarol,  Phys.Rev. {\bf D75} (2007) 055014,
  [\href{http://xxx.lanl.gov/abs/hep-ph/0612048}{{\tt hep-ph/0612048}}].

\bibitem{Gripaios:2009pe}
B.~Gripaios, A.~Pomarol, F.~Riva, and J.~Serra,  JHEP {\bf 0904} (2009) 070,
  [\href{http://xxx.lanl.gov/abs/0902.1483}{{\tt arXiv:0902.1483}}].

\bibitem{Halyo:1991pc}
E.~Halyo,  Mod.Phys.Lett. {\bf A8} (1993) 275--284.

\bibitem{Goldberger:2007zk}
W.~D. Goldberger, B.~Grinstein, and W.~Skiba,  Phys.Rev.Lett. {\bf 100} (2008)
  111802, [\href{http://xxx.lanl.gov/abs/0708.1463}{{\tt arXiv:0708.1463}}].

\bibitem{Vecchi:2010gj}
L.~Vecchi,  Phys.Rev. {\bf D82} (2010) 076009,
  [\href{http://xxx.lanl.gov/abs/1002.1721}{{\tt arXiv:1002.1721}}].

\bibitem{Campbell:2011iw}
B.~A. Campbell, J.~Ellis, and K.~A. Olive,  JHEP {\bf 1203} (2012) 026,
  [\href{http://xxx.lanl.gov/abs/1111.4495}{{\tt arXiv:1111.4495}}].

\bibitem{Matsuzaki:2012mk}
S.~Matsuzaki and K.~Yamawaki,  \href{http://xxx.lanl.gov/abs/1207.5911}{{\tt
  arXiv:1207.5911}}.

\bibitem{Chacko:2012vm}
Z.~Chacko, R.~Franceschini, and R.~K. Mishra,
  \href{http://xxx.lanl.gov/abs/1209.3259}{{\tt arXiv:1209.3259}}.

\bibitem{Bellazzini:2012vz}
B.~Bellazzini, C.~Csaki, J.~Hubisz, J.~Serra, and J.~Terning,
  \href{http://xxx.lanl.gov/abs/1209.3299}{{\tt arXiv:1209.3299}}.

\bibitem{Alonso:2012jc}
R.~Alonso, M.~Gavela, L.~Merlo, S.~Rigolin, and J.~Yepes,  JHEP {\bf 1206}
  (2012) 076, [\href{http://xxx.lanl.gov/abs/1201.1511}{{\tt
  arXiv:1201.1511}}].

\bibitem{Alonso:2012pz}
R.~Alonso, M.~Gavela, L.~Merlo, S.~Rigolin, and J.~Yepes,  Phys.Rev. {\bf D87}
  (2013) 055019, [\href{http://xxx.lanl.gov/abs/1212.3307}{{\tt
  arXiv:1212.3307}}].

\bibitem{Low:2009di}
I.~Low, R.~Rattazzi, and A.~Vichi,  JHEP {\bf 1004} (2010) 126,
  [\href{http://xxx.lanl.gov/abs/0907.5413}{{\tt arXiv:0907.5413}}].

\bibitem{Buchalla:2013rka}
G.~Buchalla, O.~Cata, and C.~Krause,
  \href{http://xxx.lanl.gov/abs/1307.5017}{{\tt arXiv:1307.5017}}.

\bibitem{Jenkins:2013sda}
E.~E. Jenkins, A.~V. Manohar, and M.~Trott,  Phys. Lett. {\bf B726} (2013) 697
  -- 702, [\href{http://xxx.lanl.gov/abs/1309.0819}{{\tt arXiv:1309.0819}}].

\end{thebibliography}
%\bibliographystyle{BiblioStyleLetter}

\providecommand{\href}[2]{#2}\begingroup\raggedright\endgroup

\end{document}